\documentclass[floatfix,twocolumn,showpacs,aps,prl,nofootinbib]{revtex4}
\usepackage{graphicx,dcolumn,amsmath,epsfig,color,colordvi,amsfonts} 
\usepackage{fancyhdr,lineno,multirow,rotating,relsize} 
\RequirePackage{xspace}

\input{babarsym}

\def\Dstp        {\ensuremath{D^{*+}}}
\def\Dstm        {\ensuremath{D^{*-}}}

\begin{document}

\title{{\large \bf \Dz - \Dzb Mixing Analyses at \babar}}
\author{Rolf Andreassen\footnote{Email:rolfa$@$slac.stanford.edu} 
(for the \babar\ collaboration)}
\affiliation{University of Cincinnati, Cincinnati, Ohio 45221, USA}
\date{31 March 2008}
\vbox{SLAC-PUB-13024 

UCHEP-07-10}

\begin{abstract}
We summarise results of analyses of $D$ meson mixing parameters
performed by the \babar\ collaboration.
\end{abstract}

\pacs{14.40.Lb, 12.15.Ff} 
\maketitle

\section{Introduction}

Understanding $D$ meson (charm) mixing is an important step in measuring \CP violation
in the charm sector. It also fills in a gap between the well-measured cases of
$K$ \cite{K:system} and $B$ \cite{B:system, Bs:system} system mixing, both of which have down-type quarks in
the intermediate state, where charm mixing has up-type quarks. 
Since mixing in the \Dz system is expected to be small in the Standard Model
\cite{CKM:GIM} (modulo the hard-to-predict effects of long-distance interactions \cite{DMixTheory}),
charm mixing also offers a chance to observe New Physics either
through \CP violation in mixing \cite{Blaylock:1995ay} or a large mass difference between the $D$ mass
eigenstates \cite{DMixTheory}. 
In this proceeding, we summarise the result of four different
approaches to measuring the $D$ mixing parameters at \babar, involving the decays
\Dz\to\Kp\pim \cite{kpi}, \Dz\to\Kp\Km or \pip\pim \cite{ll}, 
\Dz\to\Kp\pim\piz \cite{kpipi}, and \Dz\to\Kp\pim\pip\pim \cite{k3pi}.

\section{Detector}

We present analyses of \epem collisions at a center-of-mass (CM) energy of 10.58,
collected at the \babar\ detector at the \pep2 storage ring. Particle identification
is done by $dE/dx$ measurements from two tracking detectors and from measuring Cherenkov
angles in a ring-imaging detector. $D$ mesons are tagged
by reconstructing \Dstarp\to\Dz\pip and \Dstarm\to\Dzb\pim decays, and assigning flavour
according to the charge of the slow pion. 

\section{Formalism and notation}

$D$ mesons are produced in pure flavour eigenstates, $|\Dz\rangle$ or $|\Dzb\rangle$.
These flavour eigenstates are not equal to the mass and lifetime eigenstates 
\begin{eqnarray*}
|D_1\rangle &=& p|\Dz\rangle + q|\Dzb\rangle\\
|D_2\rangle &=& p|\Dz\rangle - q|\Dzb\rangle
\end{eqnarray*}
by which they propagate and decay. Therefore, a particle produced as a \Dz 
may become a \Dzb before its decay. The process is governed by the mass and
lifetime differences of the $D_1$ and $D_2$ states; these decay according to
\begin{eqnarray*}
|D_1(t)\rangle &=& e^{-i(m_1-i\Gamma_1/2)t}|D_1\rangle\\
|D_2(t)\rangle &=& e^{-i(m_2-i\Gamma_2/2)t}|D_2\rangle
\end{eqnarray*}
where $m_i,\Gamma_i$ are the mass and width of the $D_i$ state. 
We define
\begin{eqnarray*}
\Delta M &=& m_1 - m_2 \\
\Delta \Gamma &=& \Gamma_1 - \Gamma_2 \\
\Gamma &=& (\Gamma_1 + \Gamma_2) / 2 \\
x &=& \Delta M / \Gamma \\
y &=& \Delta \Gamma / 2\Gamma \\
R_M &=& (x^2+y^2)/2
\end{eqnarray*}
The quantities $x$ and $y$ are collectively referred to as mixing parameters. 
Estimates within the Standard Model vary from $10^{-4}$ (counting only short-distance effects) 
to as high as 1\%.
Establishing the presence of New Physics requires either $x$ $>>$ $y$, or 
\CP violation \cite{DMixTheory}. 

\section{Experimental approach}

The studies considered here use a common apparatus for tagging
$D$ mesons as either \Dz or \Dzb, and for measuring their decay times. 
In particular, by considering only $D$ mesons from \Dstar\to\Dz$\pi_s$,
we can use the charge of the slow pion $\pi_s$ to determine the 
production flavour of the \Dz, and measure its flight length
from the decay vertices of the \Dstar and \Dz particles.
We make use of the mass of \Dz candidates ($m_{\Dz}$) and the mass difference $\Delta m$ 
between \Dz and \Dstar candidates to extract our signal yields, and to define sidebands
for background studies. Figure \ref{kpimdm} shows distributions of these quantities
for the \Dz\to\Km\pip analysis, which may be considered typical. 

For historical reasons, $D$ mesons whose decay flavour matches their
production flavour (e.g. \Dstarp\to\Dz\pip with \Dz\to\Km\pip)
are called 'right-sign' (RS), while the opposite case is referred
to as 'wrong-sign' (WS). Wrong-sign decays may come about either through 
mixing or through doubly-Cabibbo-suppressed (DCS) Feynmann diagrams.
To distinguish the two cases, we use the decay-time distribution,
as will be shown for each decay mode.

In addition to
these two sources of wrong-sign events, there is the case where a 
correctly reconstructed \Dz is matched with a pion not from a \Dstar
decay to produce a spurious \Dstar; this is referred to as the ``mistag''
background. Another source of background is $D$ mesons reconstructed
with the correct tracks, but wrong particle assigments, or with tracks
missing; this is the ``bad \Dz'' or ``mis-reconstructed charm'' background. Finally there is 
background from combinatorics.

\section{\boldmath{\Dz \to \Kp\pim}}

\label{sectionKpi}

In the limit of small mixing and \CP conservation, the 
decay-time distribution for wrong-sign decays of mesons
produced as \Dz may be approximated as 
\begin{eqnarray}
\frac{T_{\mathrm{WS}}(t)}{e^{-\Gamma t}} &\propto& R_D + y'\sqrt{R_D}(\Gamma t) + 
 \frac{\scriptstyle 1}{\scriptstyle 4}(x'^2+ y'^2)(\Gamma t)^2
\label{kpieqn}
\end{eqnarray}
where $x'$ and $y'$ are related to $x$ and $y$ by
\begin{eqnarray*}
x' &=& x\cos\delta_{K\pi}+y\sin\delta_{K\pi} \\
y' &=& y\cos\delta_{K\pi}-x\sin\delta_{K\pi}.
\end{eqnarray*}
The angle $\delta_{K\pi}$ is the strong phase between Cabibbo-favoured (CF)
and DCS decays. The quantity $R_D$ is the amplitude, in the absence of mixing,
for the \Dz to decay by a DCS process; the term quadratic in $t$ is the amplitude,
in the absence of DCS processes, for the \Dz to mix and then decay as a \Dzb; and
the term linear in $t$ is the interference term between these two processes. 

We apply Equation \ref{kpieqn} in two ways: The first is to enforce \CP\ 
conservation by fitting both \Dz and \Dzb samples together. The second is
to search for \CP\ violation by doing two fits, calculating $x'^2$
and $y'$ for \Dz and \Dzb separately.

\begin{figure}[phtb]
  \centering
  \centerline{%
    \includegraphics[width=0.5\linewidth, clip=]{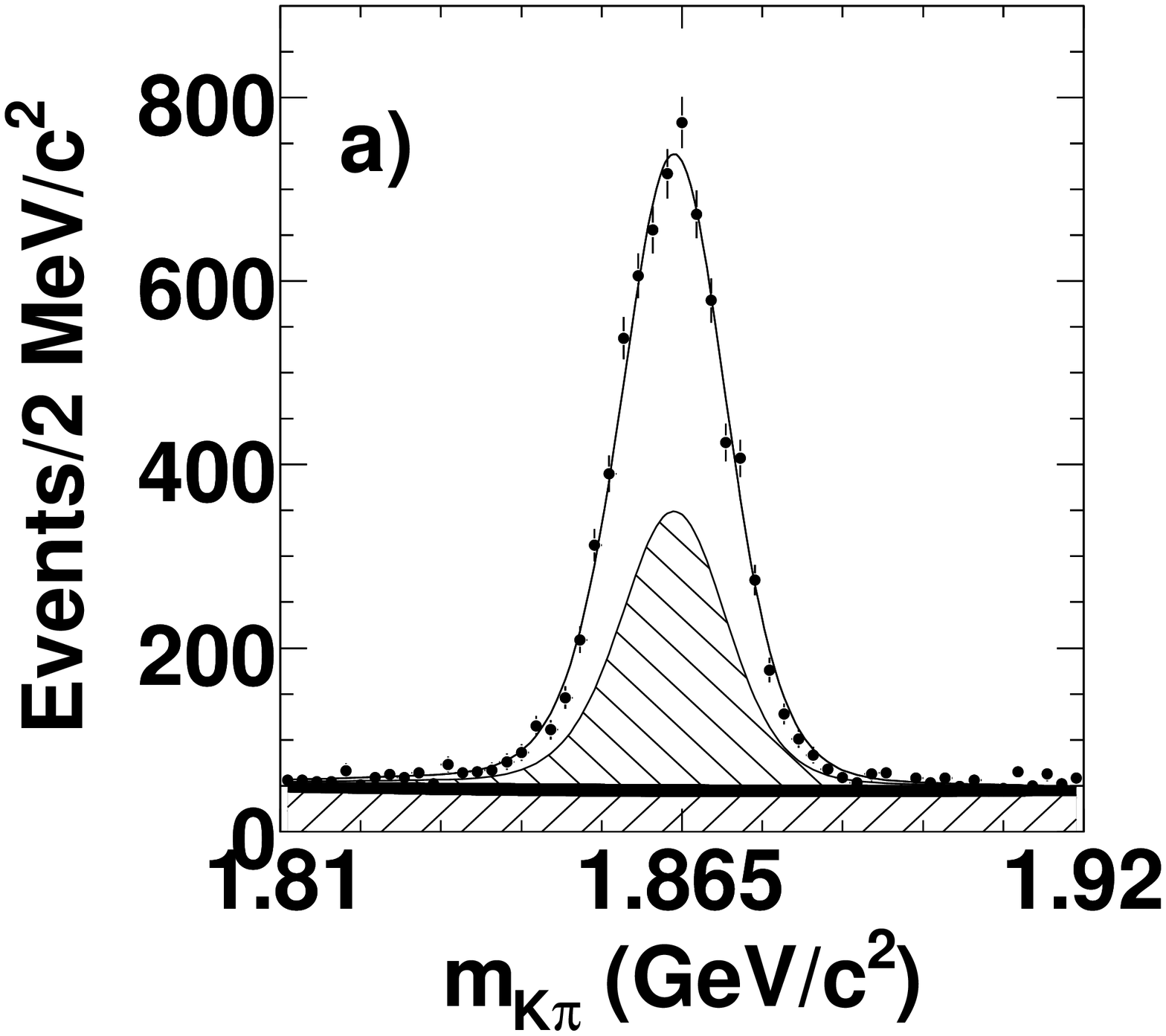}
    \includegraphics[width=0.5\linewidth, clip=]{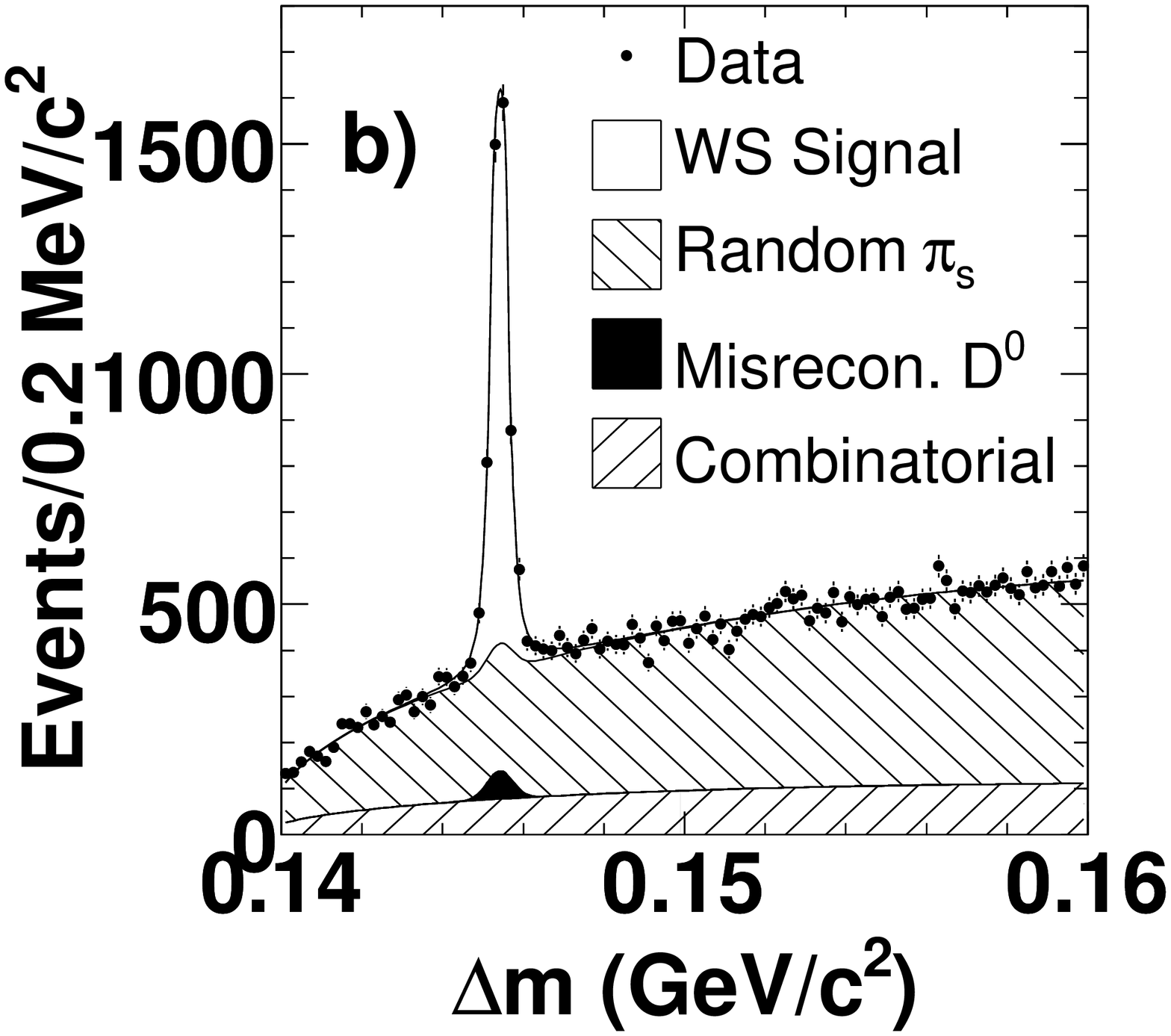}
  }
  \caption{a)  $m_{K\pi}$ for 
wrong-sign (WS) candidates with $0.1445<\Delta m<0.1465\gevcc$, and b)  $\Delta m$ for WS candidates
with $1.843<m_{K\pi}<1.883\gevcc$. The fitted PDFs are overlaid. 
}
  \label{kpimdm}
  \smallskip
\end{figure}

We use 384 fb$^{-1}$ of \epem data, pairing tracks of opposite charge
to make \Dz candidates, and then pairing these with slow pion tracks
to make \Dstar candidates. The phase space available for slow pions is 
small; we require their momentum to be greater than
0.1 \gevc in the lab frame, and less than 0.45 \gevc in the CM frame. We fit the full
decay chain, constraining the \Dstar to come from the beam spot, the \Dz and slow pion
to come from a common vertex, and the \Kmp and \pipm to come from a different common vertex.
We reject candidates
if the $\chi^2$ probability of this fit is less than 0.1\%. The \Dz decay time and
error on the decay time are taken from this fit; candidates whose decay-time error
exceeds 0.5 ps are assumed to be badly reconstructed, and thrown away, and we
also require that the decay time be between $-2$ and 4 ps.
We further require
the CM momentum of \Dz candidates to be at least 2.5 \gevc, which suppresses backgrounds
from $B$-meson decays and combinatorics. Where multiple \Dstar candidates share tracks,
we use only the candidate with the highest $\chi^2$ probability from the fit. 
With these criteria, our samples consist of
1,229,000~RS and 64,000~WS \Dz and \Dzb candidates. Figure \ref{kpimdm} shows their
distribution in $m_{K\pi}$ and $\Delta m$. 

We extract the mixing parameters using an unbinned, extended maximum-likelihood
fit, which proceeds
in three stages. The first step is to fit the $m_{K\pi}-\Delta m$ distributions
to extract shape parameters in these variables; these are then fixed in subsequent
fits. Next we fit the RS sample to extract the \Dz lifetime and resolution functions,
using the $m_{K\pi}-\Delta m$ parameters from the previous step to separate the components.
Finally we fit the WS sample for the mixing parameters using three different models. The first model
assumes no \CP\ violation and no mixing; the second permits mixing, but not \CP\
violation; the third allows both mixing and \CP\ violation. 

The $m_{K\pi}-\Delta m$ distributions are fitted to a sum of four PDFs, one each
for signal, mistags, bad \Dz and combinatorial background. Of these, the signal
peaks in both $m_{K\pi}$ and $\Delta m$. The mistagged events - correctly reconstructed
\Dz with a pion not from a \Dstar decay - peak in $m_{K\pi}$ but not in $\Delta m$. 
Bad \Dz events have a \Dz with one or more daughters missing, or assigned the wrong
particle hypothesis; they peak in $\Delta m$ but not in $m_{K\pi}$. Finally, combinatorial
background does not peak in either variable. Figure \ref{kpimdm} shows these various
shapes. The signal peak contains $1,141,500\pm 1,200$ candidates for the RS sample, and
$4,030\pm 90$ for the WS. 

\begin{figure}[phtb]
  \centering
  \centerline{%
    \includegraphics[width=0.9\linewidth, clip=]{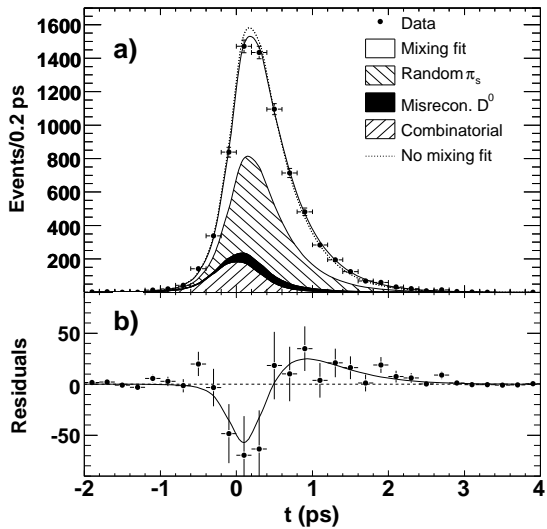}
  }
\caption{\label{kpiFitFig}a) Projections of the 
proper-time distribution of combined \Dz and \Dzb 
WS candidates and fit result integrated over the signal region
1.843 $<m_{K\pi}<$ 1.883 \gevcc and 0.1445 $<\Delta m<$ 0.1465 \gevcc.
The result of the fit allowing (not allowing) mixing 
but not \CP violation is overlaid as a solid (dashed) curve.
b) The points represent the difference 
between the data and the no-mixing fit. The solid curve
shows the difference between fits with and without mixing.
The difference between the mixing-allowed fit and the data 
is therefore the difference between the solid curve and the points,
or essentially zero.}
\end{figure}

We describe the decay-time distribution of the RS signal with an exponential convolved
with a sum of three Gaussians, whose widths are proportional to the measured event-by-event
error on the decay time. The combinatorial background is described by a sum of two
Gaussians, one of which has a power-law tail; the mistag background is described by the
same PDF as the signal, because the slow pion has little influence on the vertex fit. 
For the WS signal, we use Equation \ref{kpieqn}, convolved with the resolution function
determined by the RS fit. Figure \ref{kpiFitFig} shows the data, overlaid by these various
PDFs. From inspection, it is clear that the fit allowing mixing describes the data better
than the one which imposes zero mixing. 

\begin{figure}[phtb]
  \centering
  \centerline{%
    \includegraphics[width=0.95\linewidth, clip=]{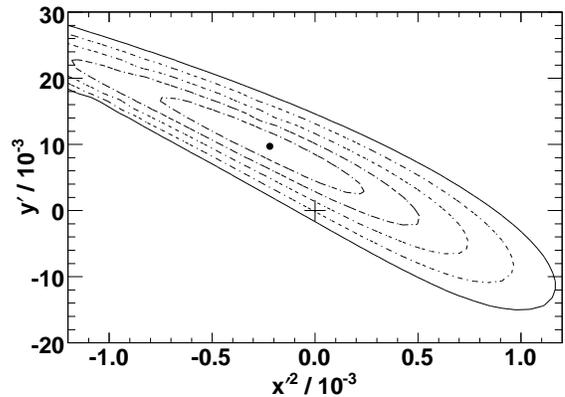}
}
\caption{The central value (point) and confidence-level (CL) contours for 
$1-\mbox{CL}=0.317\ (1\sigma)$, $4.55\times10^{-2}\ (2\sigma)$, 
$2.70\times10^{-3}\ (3\sigma)$, $6.33\times10^{-5}\ (4\sigma)$ and
$5.73\times10^{-7}\ (5\sigma)$, calculated from the change in the value
of $-2\ln{\cal L}$ compared with its value at the minimum.
Systematic uncertainties are included. The no-mixing point is shown 
as a plus sign~($+$).}
\label{kpiContour}
\end{figure}

Figure \ref{kpiContour} shows the likelihood contours of the mixing parameters from 
the fit allowing mixing but not \CP\ violation, including systematic uncertainties. 
The point of maximum probability is in the unphysical region where $x'^2$ is negative;
adjusting for this by moving to the most likely point in the physical region,
$x'^2=0$, $y'=6.4 \times 10^{-3}$, we find that $-2\Delta\ln{\mathcal{L}}$ is 23.2 units
between the most likely physical point and the point of no mixing. Including the
systematic uncertainties, we thus find mixing at a significance of 3.9 $\sigma$.
Table \ref{kpiresults} shows the results of our fits in more detail; we find no evidence
for \CP\ violation, as shown by the asymmetry $A_D = (R_D^{+} - R_D^{-})/(R_D^{+} + R_D^{-})$
where subscript '$+$' indicates only the \Dz sample was used, and '$-$' indicates the
\Dzb sample.

\begin{table}[bth]
  \caption{Results from the different fits.
  The first uncertainty listed is statistical and the second systematic.}
  \label{kpiresults}
  \centering\small
  \begin{ruledtabular}
    \begin{tabular}{lcr@{~$\pm$}r@{~$\pm$}r}
     Fit type & Parameter & \multicolumn{3}{c}{Fit Results ($/10^{-3}$)}  \\
    \hline
    No \CP viol. or mixing & $R_D$ & $3.53 $ & $ 0.08 $ & $ 0.04$\\
    \hline
    \multirow{3}{1.7cm}{No \CP\\ violation}
    &  $R_D$     & $3.03$ & $0.16$ & $ 0.10$ \\
    &  $x'^2$    & $-0.22$ & $0.30$ & $ 0.21$   \\
    &  $y'$      & $9.7$ & $4.4$ & $ 3.1$      \\
   \hline
    \multirow{5}{1.7cm}{\CP\\ violation \\ allowed}
    & $R_D$      & $3.03$ &$0.16$ & $0.10$  \\
    & $A_D$      & $-21$ & $52$ & $15$  \\
    & $x'^{2+}$  & $-0.24 $ & $ 0.43 $ & $ 0.30 $\\
    & $y'^+$     & $ 9.8  $ & $ 6.4  $ & $ 4.5  $\\
    & $x'^{2-}$  & $-0.20 $ & $ 0.41 $ & $ 0.29 $\\
    & $y'^-$     & $ 9.6  $ & $ 6.1  $ & $ 4.3  $
  \end{tabular}
  \end{ruledtabular}
\end{table}

We evaluate systematic uncertainties from three sources: Variations
in the fit model, in the selection criteria, and in our procedure for
dealing with track-sharing \Dstar candidates. The most significant source
of systematic uncertainty in $R_D$ and the mixing parameters
is from the fit model for the long-lived background
component caused by other $D$ decays in the signal region, followed by the
presence of a non-zero mean in the time-resolution function, caused by 
small misalignments in the detector. For the asymmetry $A_D$, the dominant
contribution is uncertainty in modeling the 
differences between \Kp and \Km absorption in the detector.

\section{\boldmath{\Dz \to \KpKm} or \boldmath{\pip\pim}}

For $D$ mesons decaying to \CP eigenstates, mixing changes the decay time distribution
in such a way that we may, to a good approximation, consider the decays exponential
with changed lifetimes (\cite{Bergmann:2000id})
\begin{eqnarray*}
\tau^+ &=& \tau^0\left[1+|q/p|\left(y\cos\phi_f - x\sin\phi_f\right)\right]^{-1} \\
\tau^- &=& \tau^0\left[1+|p/q|\left(y\cos\phi_f + x\sin\phi_f\right)\right]^{-1}
\end{eqnarray*}
where $\tau^0$ is the lifetime for decays to final states which are not \CP eigenstates,
and $\tau^+$ ($\tau^-$) is the lifetime for \Dz (\Dzb) decays to \CP-even states. We
can combine the three lifetimes into quantities
\begin{eqnarray*}
y_{\CP} &=& \tau^0/\langle\tau\rangle - 1 \\
\Delta Y &=& \left(\tau^0A_\tau\right)/\langle\tau\rangle.
\end{eqnarray*}
Here $\phi_f$ is the \CP-violating phase $\phi_f=\arg(q\overline A_f / pA_f)$, 
$A_f$ ($\overline A_f$) being the amplitude for \Dz (\Dzb) decaying to the final state
$f$. $\langle\tau\rangle$ is the average of $\tau^+$ and $\tau^-$, 
and $A_\tau$ is their asymmetry $\left(\tau^+-\tau^-\right)/\left(\tau^++\tau^-\right)$.
In the absence of mixing, both $y_{\CP}$ and $\Delta Y$ are zero. In the absence
of \CP violation in the interference of mixing and decay (ie, $\phi_f=0$),
$\Delta Y$ is zero and $y_{\CP}=y$. 

For this analysis, we use 384 fb$^{-1}$ of \babar\ data, and measure the lifetimes for the \CP-even decays\footnote{Charge conjugation
is implied throughout unless otherwise noted.} \Dz\to\Kp\Km and \Dz\to\pip\pim,
and for \Dz\to\Km\pip, which is not a \CP eigenstate and thus gives our $\tau^0$. 

In addition to particle identification requirements,
the cosine of the helicity angle (defined as the angle between the momentum of the positively charged \Dz daughter
in the \Dz rest frame, and the \Dz's momentum in the lab frame) is required less than 0.7;
this suppresses combinatorial backgrounds. \Dz candidates are then combined with pions to produce \Dstar candidates.
Electrons are rejected by combining pion candidates with each other track in the event
and vetoing those which form a good photon conversion or pion Dalitz decay, as well
as by \dedx measurements. The requirements for slow pions and the vertex fit of the \Dstar
are the same as for the \hbox{\Dz\to\Km\pip} analysis (Section 5). 

\begin{figure}[!ht]
\hbox to \hsize{
 \includegraphics[width=0.45\linewidth]{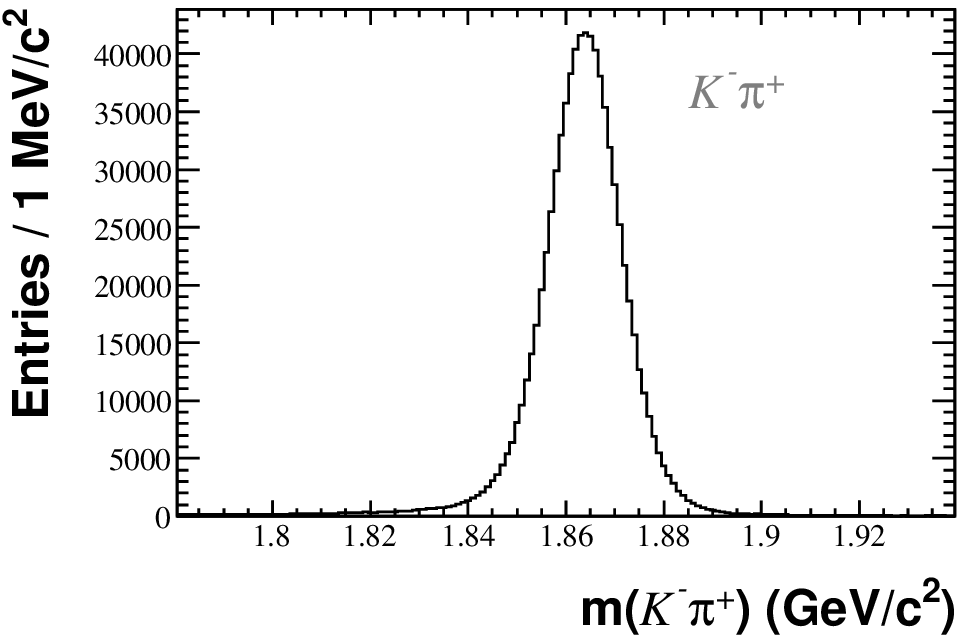}\hfill  
 \includegraphics[width=0.45\linewidth]{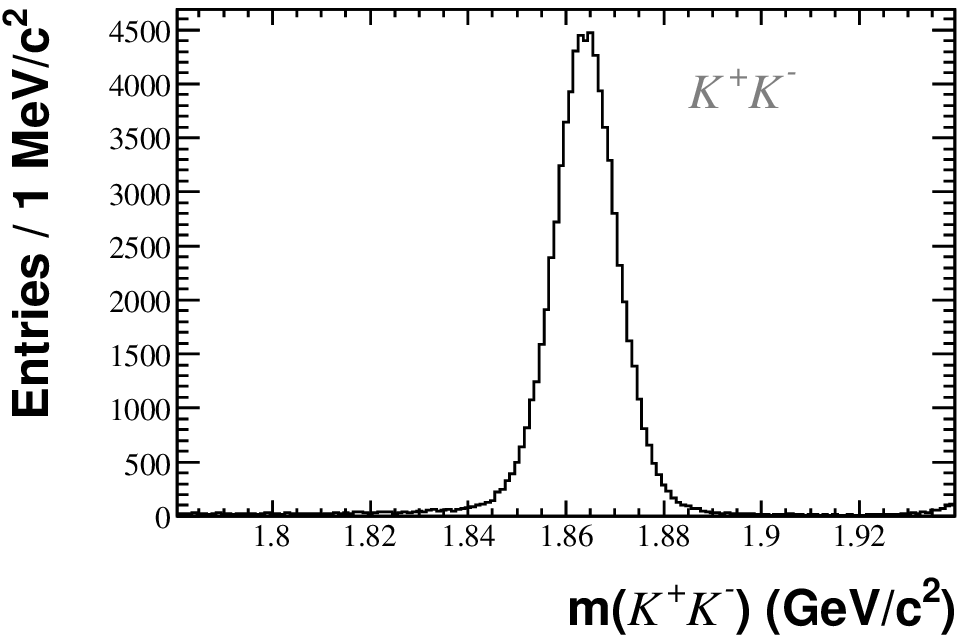}
}
\hbox to \hsize{
 \includegraphics[width=0.45\linewidth]{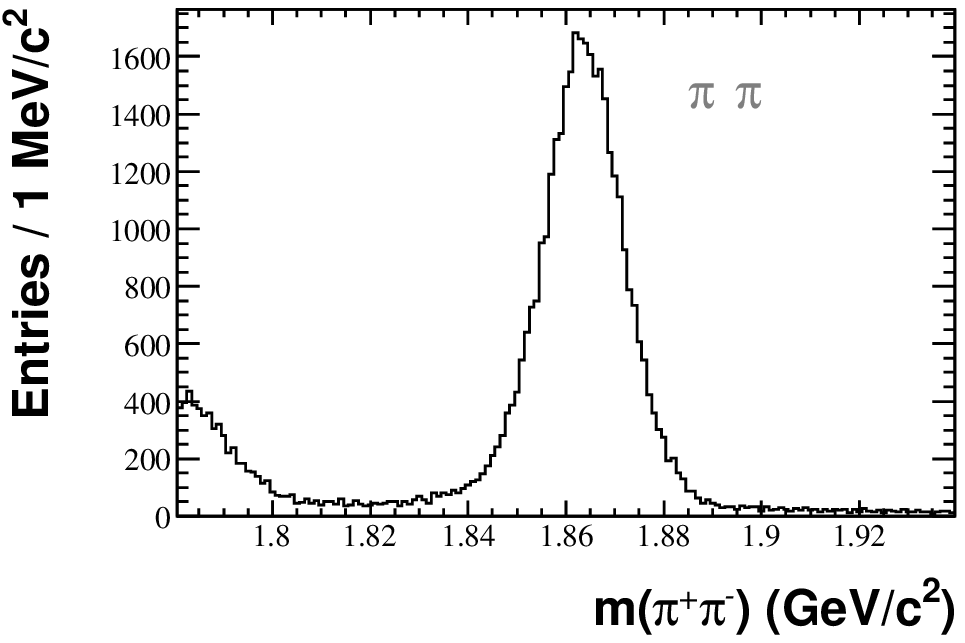}\hfill
}
\caption{\label{kkd0massfig}Reconstructed $D^0$ mass distributions
for the three $D^0$ samples, within $\pm$0.8~\mevcc of the $\Delta
m$ peak.}
\end{figure}


Figure \ref{kkd0massfig} shows the mass distributions of \Dz candidates; Table \ref{yieldtable}
shows the yield and purity of the samples, calculated using events within a 15 \mevcc \Dz mass 
and 0.8 \mevcc $\Delta m$ window. We fit the decay time distributions of these samples using
an unbinned maximum likelihood fit to all five decay modes simultaneously, using separate PDFs
for signal decays, mistagged events, mis-reconstructed charm events, and combinatorial background. 

\begin{table}[t]
\begin{tabular}{lcr}\hline\hline
Sample &  Size & Purity (\%) \\
\hline
$K^-\pi^+$         & 730,880 & 99.9  \\
$K^-K^+$           &  69,696 & 99.6  \\
$\pi^-\pi^+$       &  30,679 & 98.0  \\
\hline\hline
\end{tabular}
\caption{\label{yieldtable} Sample sizes and purities.}
\end{table}

As with the \Dz\to\Km\pip study, we model the decay-time distribution of signal events using a simple exponential
convolved with a sum of three Gaussians for the resolution. Each Gaussian has
a width proportional to the event-by-event error on the measured decay time; their
mean is common, and allowed to be offset from zero to account for any effects of detector
mis-alignment. Mis-tagged events - that is, events with a correctly reconstructed
\Dz, but wrongly assigned slow pion - account for about 0.4\% of the sample; of these,
half will have the wrong flavour assignment to the \Dz. However, they have the same
decay-time distribution and resolution as true signal. Hence we model these events
using the signal PDF, but reversing the flavour assignment. 

Mis-reconstructed charm events have an exponential decay-time distribution, which 
we convolve with a single Gaussian. The fraction of such events is obtained from simulation,
which we check by comparing data and Monte Carlo in the sidebands 
$1.89<m_{\Dz}<1.92\gevcc$ and $0.151<\Delta m<0.159\gevcc$.
We estimate the charm background as $(0.009\pm 0.002)\%$ of events in the signal region for \Dz\to\Km\pip,
$(0.2\pm 0.1)\%$ for \Dz\to\Kp\Km, and $(0.15\pm 0.15)\%$ for \Dz\to\pip\pim. For
combinatorial background, we model
the decay-time distribution as the sum of a Gaussian and a modified Gaussian with a power-law
tail, the latter accounting for a long-lived component. Each decay mode has its own shape
for combinatorial background, the shapes being determined from fits to the sideband
regions; the fraction of this background is again estimated from Monte Carlo with uncertainties
derived from comparison of MC and data. We find  
$(0.032\pm 0.003)\%$ in the \Dz\to\Km\pip mode, $(0.16\pm 0.02)\%$ in \Dz\to\Kp\Km, and
$(1.8\pm 0.2)\%$ in \Dz\to\pip\pim. 

We consider several sources of systematic error, including
variations of the signal and background models, changes to the event
selection, and detector effects. We vary the models by changing the
signal PDF shape and size, as well as the position of the signal box. We
also test our resolution model by forcing the common mean of the three Gaussians
to zero, and by allowing it to float separately for different bins of the
\Dz polar angle. Of these effects, the largest systematic uncertainty
derives from widening the \Dz mass window, which increases the amount
of badly-reconstructed signal events in the sample. 

We vary the mis-reconstructed charm model by changing its fraction
in the fit, by varying its effective lifetime, by using a different
sideband region, and by using a decay time distribution obtained from 
Monte Carlo instead of the sideband data. Due to the purity of the data,
these effects are all small, the largest being from varying the background
fraction in the \Dz\to\pip\pim mode, where the purity is worst. 

We vary our event selection criteria in two ways: By throwing out or keeping
all multiple candidates (as opposed to selecting the candidate with the best
$\chi^2$ probability for its vertex fit), and by changing the acceptable
range of errors on decay times. The last, which changes the amount of
poorly reconstructed signal events, has the largest effect.

Finally, we consider effects of our understanding of the detector by 
repeating our analysis with different misalignment parameters. This changes
our fitted lifetimes by up to 3 fs; but since the lifetimes change by similar
amounts, and we are considering ratios of lifetimes, the effect on the mixing
parameters is small. All these systematic effects are summarised in Table \ref{llSystTable}.

\begin{linenomath}
\begin{table}[t]
  \caption{Summary of systematic uncertainties on $y_{\CP}$ and $\Delta Y$, separately for $KK$ and $\pipi$
and averaged over the two \CP modes, in percent.}
  \label{llSystTable}
  \begin{center}
  \begin{tabular}{l|cccccccc}
    \hline
    \hline
&  &\multicolumn{3}{c}{$\sigma_{y_{\CP}}$ (\%)}& &\multicolumn{3}{c}{$\sigma_{\Delta Y}$ (\%)}\\

\cline{3-5}\cline{7-9}
Systematic & & $KK$ & $\pipi$ & Avg. & &$KK$ & $\pipi$ & Avg. \\\hline
Signal model          & &  0.130& 0.059     & 0.085     & &   0.072    &   0.265    &   0.062    \\
Charm bkg             & &  0.062& 0.037     & 0.043     & &   0.001    &   0.002    &   0.001    \\
Comb. bkg             & &  0.019& 0.142     & 0.045     & &   0.001    &   0.005    &   0.002    \\
Selection criteria    & &  0.068& 0.178     & 0.046     & &   0.083    &   0.172    &   0.011    \\
Detector model        & &  0.064& 0.080     & 0.064     & &   0.054    &   0.040    &   0.054    \\\hline
Quadrature sum        & &  0.172& 0.251     & 0.132     & &   0.122    &   0.318    &   0.083    \\\hline\hline
  \end{tabular}
  \end{center}
\end{table}
\end{linenomath}

\begin{table}[b]
\begin{tabular}{l|r}
\hline\hline
Mode & Lifetime (fs) \\ \hline
\Dz\to\Km\pip            & 409.33 $\pm$ 0.70 \\
\Dz (\Dstp) \to \Kp\Km   & 401.28 $\pm$ 2.47 \\
\Dz (\Dstm) \to \Kp\Km   & 404.47 $\pm$ 2.52 \\
\Dz (\Dstp) \to \pip\pim & 407.64 $\pm$ 3.68 \\
\Dz (\Dstm) \to \pip\pim & 407.26 $\pm$ 3.73\\
\hline\hline
\end{tabular}
\caption{\label{llLifetimesTable} Measured lifetimes for the different decay modes. Uncertainties
are statistical only.}
\end{table}

The results of these decay-time fits are shown in Table~\ref{llLifetimesTable}. From the measured
lifetimes, we extract 
\begin{eqnarray*}
y_{\CP} &=& 1.24\pm 0.39\stat \pm 0.13\syst]\% \\
\Delta Y &=& [-0.26\pm 0.36\stat \pm 0.08\syst]\%
\end{eqnarray*}
which is evidence for \Dz-\Dzb mixing at the 3-sigma level, 
and consistent with \CP\ conservation. This amount of \Dz-\Dzb mixing
is consistent with Standard Model predictions. 

\section{\boldmath{\Dz \to \Kp\pim\piz}}

For the case of \Dz decays to three-body final states, we can modify
Equation \ref{kpieqn} to give a decay-time distribution for each point
in the decay phase space:
\begin{eqnarray}
\label{kpipieqn} 
\mathcal{A}(P,t) &=& e^{-\Gamma t}\Big[|\overline{A_P}|^2 \\
\nonumber&&+ |\overline{A_P}||A_P|\left(y''\cos\delta_P-x''\sin\delta_P\right)\Gamma t \\
\nonumber &&+ |A_P|^2(x''^2+y''^2)(\Gamma t)^2\Big].
\end{eqnarray}
In analogy with Equation \ref{kpieqn}, $\overline {A_P}$ is the amplitude (in the absence
of mixing) for \Dz mesons to decay by a DCS process to the point $P$  on the Dalitz plot.
The term quadratic in time is the amplitude (in the absence of DCS processes) for the \Dz
to mix before its decay, and then decay to the point $D$ by a CF process. Within this term,
the factor $A_P$ is the amplitude for the CF decay, while the remaining factors are the
mixing amplitude. The term
linear in time is the interference between the DCS and mixing terms. The quantity $\delta_P$ is the phase
of the intermediate states in the decay, relative to some reference resonance. 
As with the \Dz\to\Km\pip case, 
an unknown strong phase $\delta_{K\pi\pi^0}$ between CF and DCS decays prevents us measuring $x$ and $y$
directly; instead we are sensitive to
\begin{eqnarray*}
x'' &=& x\cos\delta_{K\pi\pi^0}+y\sin\delta_{K\pi\pi^0} \\
y'' &=& y\cos\delta_{K\pi\pi^0}-x\sin\delta_{K\pi\pi^0}.
\end{eqnarray*}

As with the previous two analyses, we use 384 fb$^{-1}$ of \babar\ data, reconstructing
\Dz\to\Km\pip\piz candidates from two oppositely-charged tracks and two photon candidates
with energy at least 100 \mev. The \piz candidate is required to have a lab momentum of at least
350 \mevc, and a mass-constrained fit probability of at least 1\%. The slow pion is required
to have a momentum transverse to the beam axis of at least 120 \mevc, and the \Dz candidate
to have a CM momentum of at least 2.4 \gevc. As in the previous two analyses, we extract 
the \Dz decay time, with error, from a vertex fit constraining the \Dstar to the beam spot;
this fit is required to have a $\chi^2$ probability of at least 1\%. 

\begin{figure}[!tb]
\begin{center}
\includegraphics[width=\linewidth]{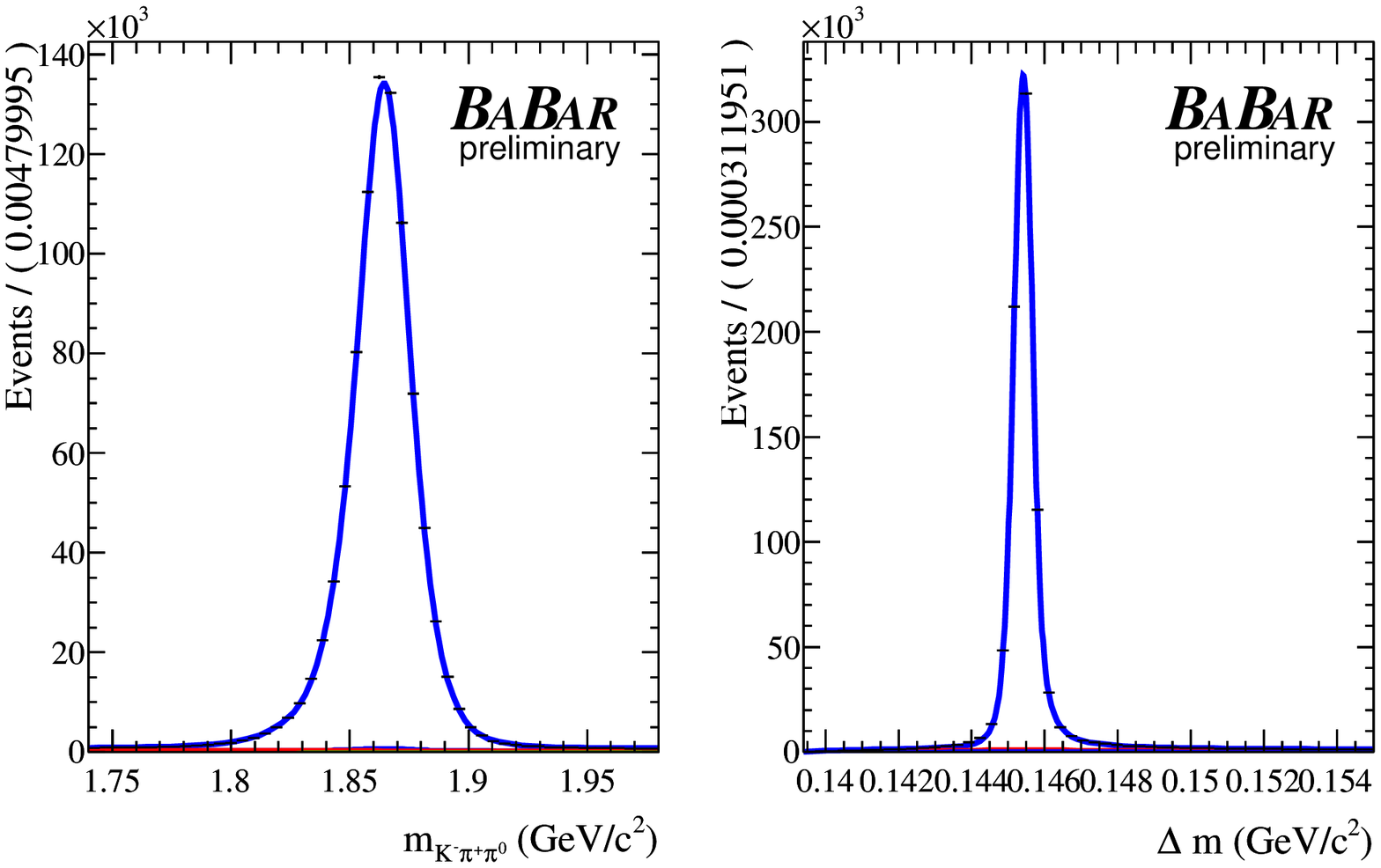}
\includegraphics[width=\linewidth]{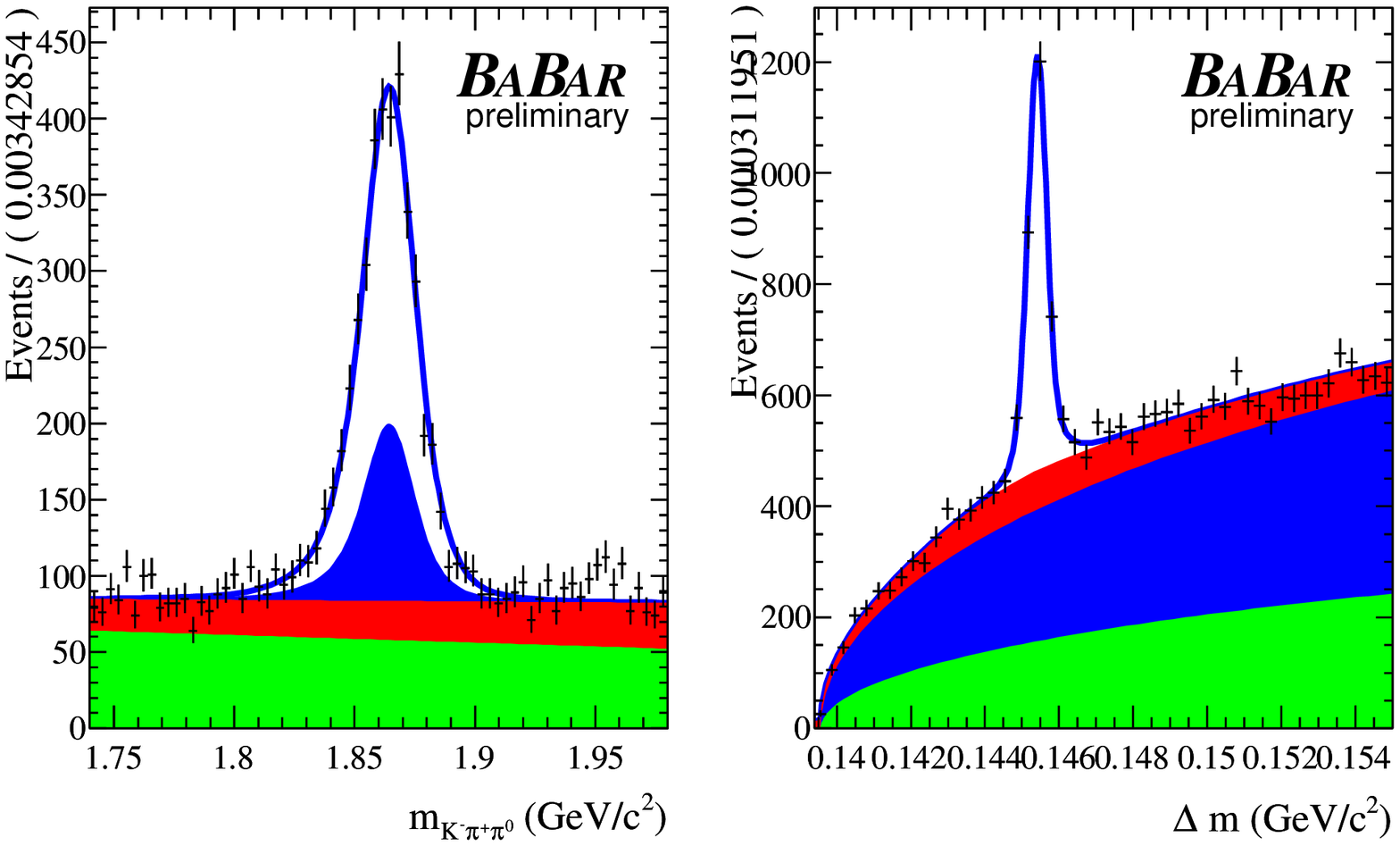}
\caption{
 (Distributions of RS  (top) and WS (bottom) data (points with error bars)
with fitted PDFs (dashed line) overlaid.  The $m_{K\pi\pi}$ distribution (left) requires
$0.145 < \Delta m < 0.146\gevcc$; the $\Delta m$ distribution (right) requires
$1.85 < m_{K\pi\pi} < 1.88gevcc$.
 The white regions represent
signal events, the light gray (blue)  misassociated
$\pi_s^{\pm}$ events, the medium gray (red)
correctly associated $\pi_s^{\pm}$ with misreconstructed
\Dz\ events, and the dark gray  (green) remaining combinatorial background.
\vspace{-5ex}}
\label{kpipidmd}
\end{center}
\end{figure}

Figure \ref{kpipidmd}
shows the $m_{K\pi\pi}$ and $\Delta m$ distributions that result from these
criteria. We fit these distributions as described for the \Dz\to\Km\pip study
in Section 5; the fit to the WS sample uses shape parameters
from the RS fit, suppressing the associated systematics. Table \ref{kpipiYields} shows
the yields for each component. 

\begin{table}[htbp]
\begin{center}
\begin{tabular}{|l|c|c|}
\hline
Category      &  N events (RS)   &   N events (WS) \\
\hline
Signal        & $639802\pm 1538$        & $1483\pm 56$ \\
Combinatoric  & $1537\pm 57$       &   $499\pm 57$       \\
Mistag        & $2384\pm 57$       &       $765\pm 29$   \\
Bad \Dz       & $3117\pm 93$       &    $227\pm 75$      \\
\hline
\end{tabular}
\end{center}
\caption{\label{kpipiYields}
Number of RS and  WS events of signal and background in the $m_{D^0}$ and $\Delta m$ signal region.}
\end{table}

We compute the quantity $A_P$ in Equation \ref{kpipieqn}, the time-independent amplitude of 
CF decays to the point $P$ on the Dalitz plot, by fitting the RS Dalitz plot
to an isobar model, using the signal and background fractions obtained in the fit
to the $m_{K\pi\pi}-\Delta m$ distribution. The background PDF is empirically
determined from the $m_{K\pi\pi}-\Delta m$ sidebands, and its fraction is set
to the background fraction derived from the $m_{K\pi\pi}-\Delta m$ fit. 

\begin{figure}[!tb]
\begin{center}
\includegraphics[width=\linewidth]{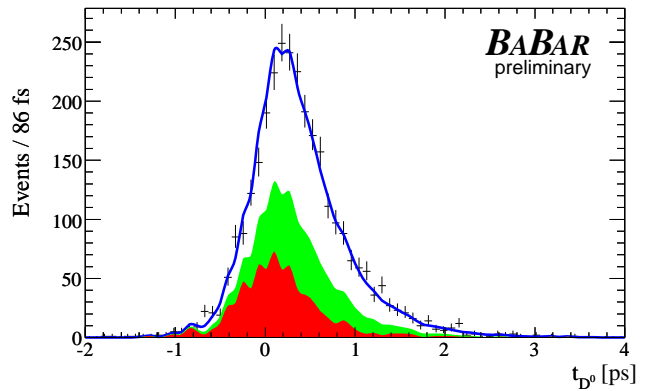}
\caption{\label{kpipiFitFig} WS \Dz decay-time distribution (crosses)
with fit (solid blue line) overlaid. The green and red regions show the
mistag and (combinatoric+bad-\Dz) backgrounds respectively. These backgrounds
are taken from sideband data, which accounts for their jagged shape. 
\vspace{-5ex}}
\end{center}
\end{figure}

With $A_P$ (or more accurately, the phases and amplitudes for intermediate
resonances from which $A_P$ can be calculated) known, we then go on
to fit the WS sample simultaneously to the Dalitz plot. We thereby determine $\overline {A_P}$,
and the decay-time distribution, to extract the mixing parameters. The signal decay-time
PDF is taken as Equation \ref{kpipieqn} convolved with a sum of three Gaussians,
as described for the previous two analyses; the parameters of the Gaussians are
extracted from a fit to the RS decay-time distribution, and fixed in the WS fit. 
For the background components, mistagged events are described by the RS parameters,
since they contain correctly reconstructed \Dz mesons; the other two background components
are described empirically using the sidebands. Figure \ref{kpipiFitFig} shows the
WS fit projected to the decay time. From this we extract 
$x'' =$  2.39  $\pm$ 0.61   (stat.) $\pm$ 0.32 (syst.)\% and $y'' =$  -0.14 $\pm$  0.60  (stat.) $\pm$ 0.40 (syst.) \%.
This excludes the no-mixing hypothesis at the 99\% confidence level. 

\section{\boldmath{\Dz\to\Kp\pim\pip\pim}}

As in the case of \Dz\to\Km\pip\piz, the four-body final state
\Kp\pim\pip\pim has a decay-time distribution which varies across
the phase space. However, since the phase space is enlarged by one
dimension, and the data sample for this analysis is smaller, we do
not fit for a decay time at each phase-space point. Instead we 
integrate across phase space to get a WS to RS decay-rate ratio of 
\begin{eqnarray}
  \label{k3pieqn}
  \frac{\Gamma_{\mathrm{WS}}(t)}{\Gamma_{\mathrm{RS}}(t)}
  &=& \tilde R_D + 
  \alpha\tilde y'\sqrt{\tilde R_D}(\Gamma t)  \\
  \nonumber && + \frac{\scriptstyle 1}{\scriptstyle 4}(\tilde x'^2+\tilde y'^2)(\Gamma t)^2
\end{eqnarray}
where a tilde indicates integration over phase space. The quantity $\alpha$ is a 
suppression factor accounting for strong-phase variation across phase space; in
effect it measures the amount of information we lose by the integration procedure. 
As in the \Dz\to\Km\pip analysis, $\tilde R_D$ is the amplitude for doubly-Cabibbo-suppressed
decays, the term quadratic in time is the amplitude for mixed decays, and the term 
linear in time is the interference between the two. Again we account for an unknown
strong phase by using variables
\begin{eqnarray*}
  \tilde x' &=& x\cos\tilde\delta + y\sin\tilde\delta \\
  \tilde y' &=& y\cos\tilde\delta - x\sin\tilde\delta
\end{eqnarray*}
where $\tilde\delta$ is the strong phase difference integrated across phase
space. Equation \ref{k3pieqn} assumes \CP conservation. To account for possible
\CP violation in interference between DCS and mixed contributions, we introduce 
the integrated \CP-violation phase $\tilde\phi$, and parametrise \CP violation
in the mixing itself with $|p/q|$. This allows us to make the substitutions
\begin{eqnarray*}
  \alpha\tilde y &\to & |p/q|^{\pm 1}\left(\alpha\tilde y' \cos\tilde\phi\pm \beta\tilde x' \sin\tilde\phi\right) \\
  \left(x^2 + y^2\right) &\to & |p/q|^{\pm 2} \left(x^2 + y^2\right) 
\end{eqnarray*}
in Equation \ref{k3pieqn}, applying plus signs for the \Dz sample and minus signs
for \Dzb. $\beta$ is an information-loss parameter analogous to $\alpha$, in this
case accounting for phase-space variation in $\phi$. 

This analysis uses a 230.4 fb$^{-1}$ \babar\ dataset. The reconstruction procedure
is analogous to that of the previous three analyses, the main difference being
the requirement that neither pion pair have an invariant mass within 20 \mevcc
of the $K_S^0$ mass of 0.4977 \gevcc. We demand a \Dz CM momentum requirement of at least 
2.4 \gevc. Two vertex fits are performed, one for the \Dz candidate, required to have
a $\chi^2$ probability of at least 0.5\%, and one for the full \Dstar decay tree.
For the latter, from which we derive our decay-time value and error,
the \Dstar is constrained to come from the beam-spot, and the probability is required
to be at least 1\%. The mean $\sigma_t$ for signal events is 0.29 ps; events with
$\sigma_t>0.5$ ps are rejected. The signal yields are calculated from a fit
to the ($m_{K3\pi}, \Delta m$) distribution; Table \ref{k3piyields} shows the results.

\begin{table}[!htp]
\caption{\label{k3piyields}
Signal yields determined by the
two-dimensional fit to the $(m_{K3\pi},\Delta m)$
distributions for the WS and RS samples.  
Uncertainties are calculated from the fit.}
\begin{center}
\begin{tabular}{lcc}
\hline
        & \Dz\                              &   \Dzb\ \\
\hline
 WS     & $(1.162  \pm 0.053) \times 10^3$  &   $(1.040  \pm 0.051) \times 10^3$ \\
 RS     & $(3.511  \pm 0.006) \times 10^5$  &   $(3.492  \pm 0.006) \times 10^5$ \\
 \hline
\end{tabular}
\end{center}
\end{table}

\begin{figure}
\begin{center}
\includegraphics[width=0.98\linewidth, trim=6cm 0pt 0pt 5.8cm, clip]{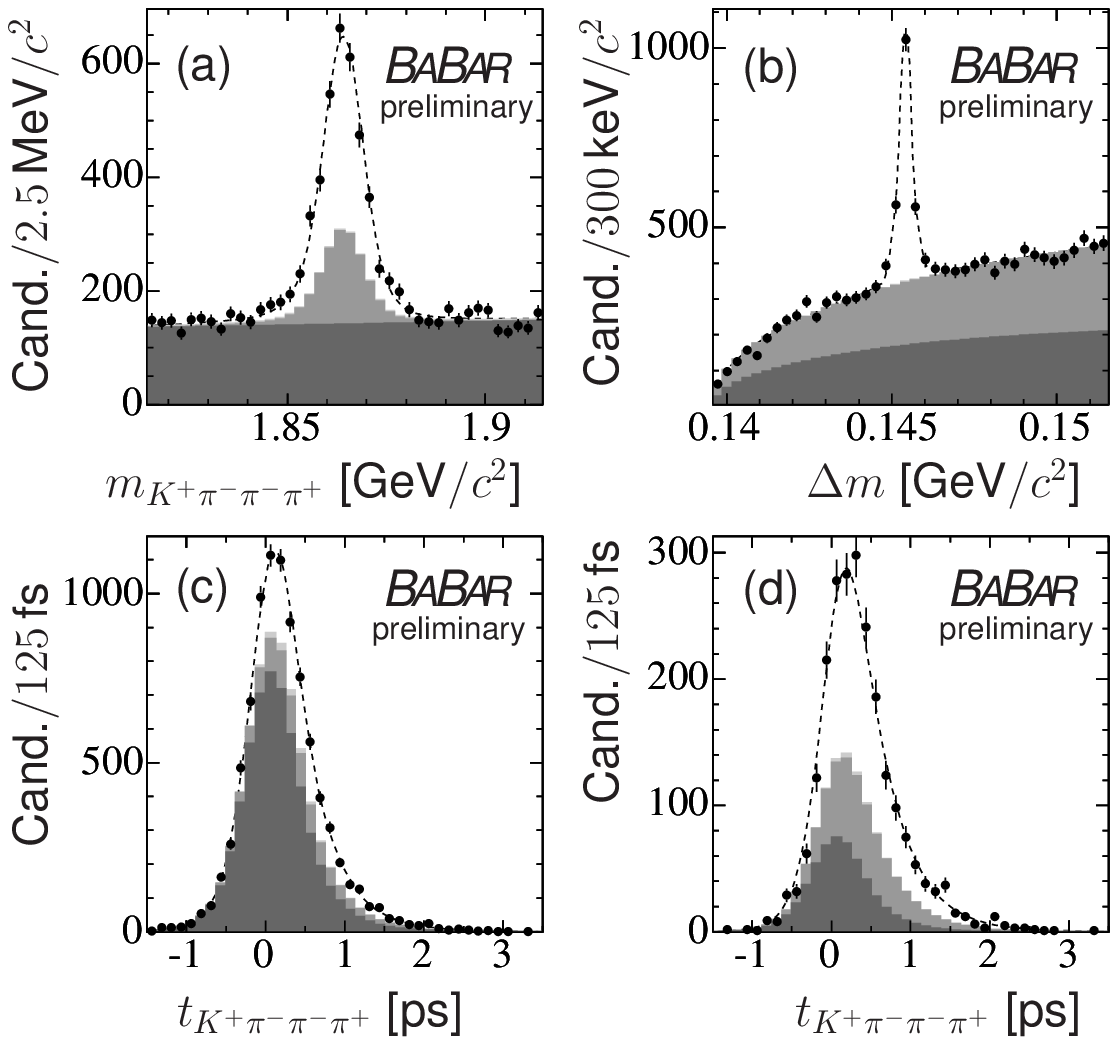}
\caption{\label{k3piFitFig}
Distributions of WS data with fitted PDF overlaid.
The light gray shows mis-reconstructed charm, 
the medium gray shows mistagged events, and
the dark gray shows combinatorial background.}
\end{center}
\end{figure}

\begin{figure}
\begin{center}

\begin{tabular}{p{0.48\linewidth}p{0.48\linewidth}}
\includegraphics[width=\linewidth]{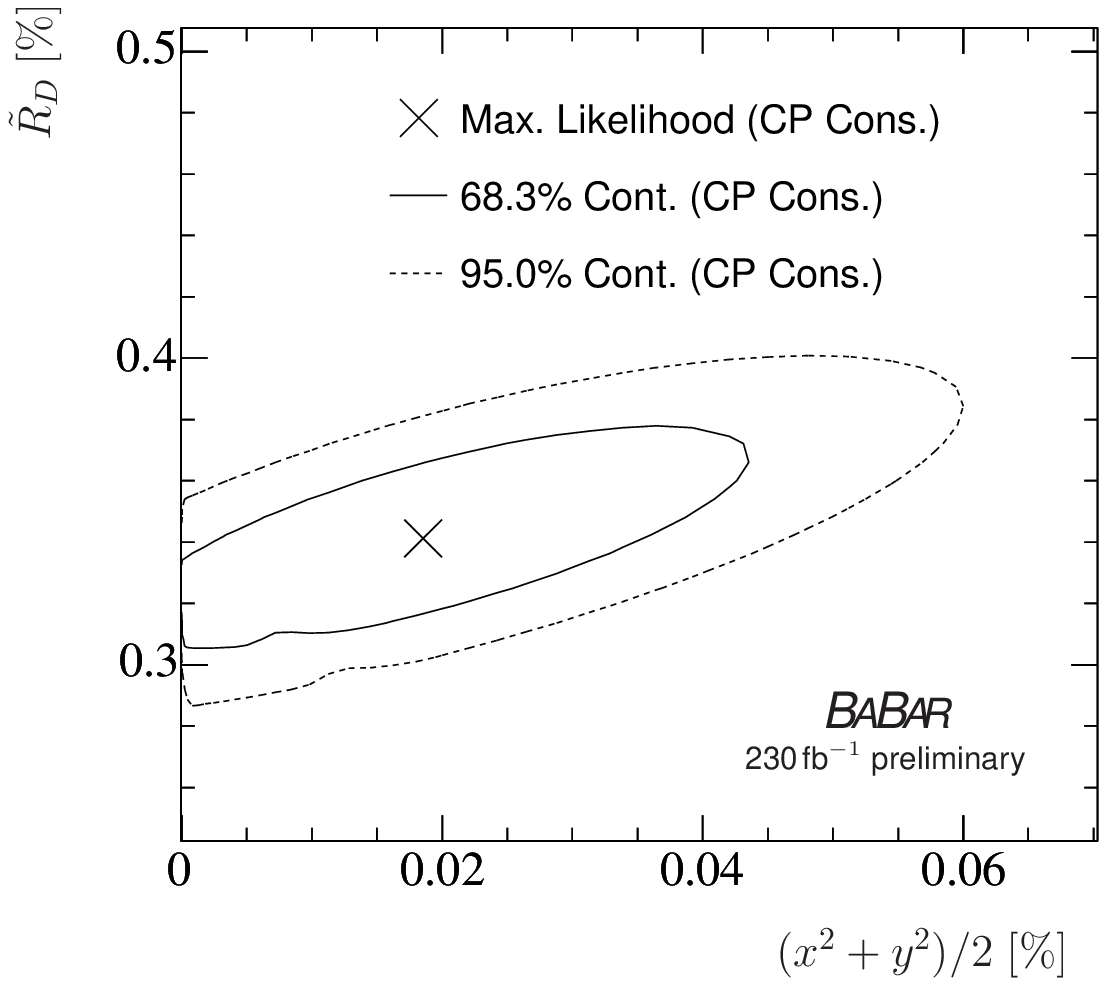}
&
\includegraphics[width=\linewidth]{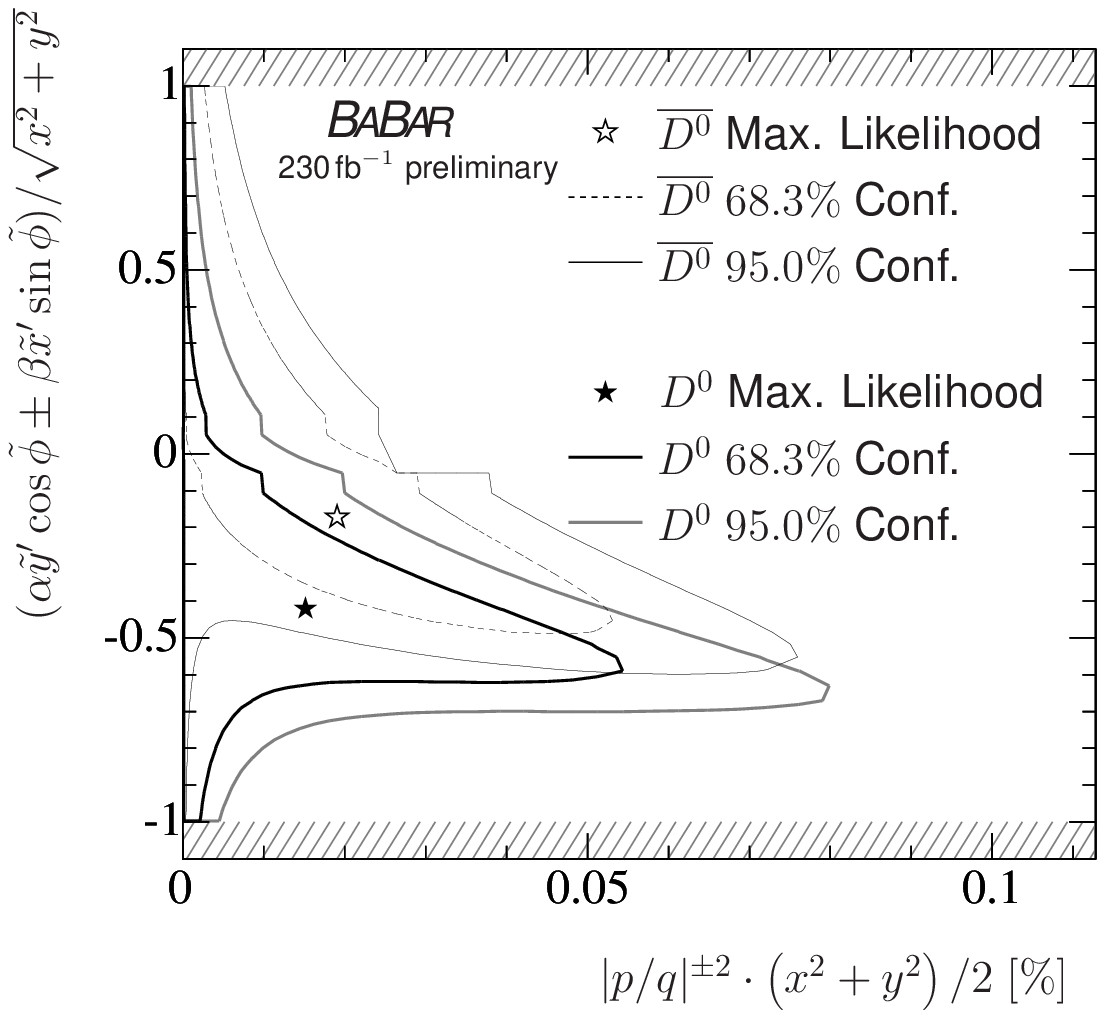}
\end{tabular}
\caption{\label{k3picont}
Left: Contours of constant $\Delta\ln\mathcal{L}=1.15,3.0$ in terms of
the doubly Cabibbo-suppressed amplitude and the time-integrated
mixing rate.  Right: Contours of constant $\Delta\ln\mathcal{L}=1.15,3.0$ in terms of
the normalized interference term and the integrated
mixing rate, for the \Dz\ and \Dzb\ samples separately.
The hatched regions are physically forbidden.
}
\end{center}
\end{figure}

The ($m_{K3\pi}, \Delta m$) fit which extracts the signal yields also determines
shape parameters for those two variables; these are then used in a three-dimensional
fit which also includes the time distribution. The decay time function for RS events
is a simple exponential convolved with a double Gaussian, with widths proportional
to $\sigma_t$ and separate means. For mistagged events we use the RS decay-time PDF;
for mis-reconstructed \Dz component we use the signal PDF; and for combinatorial background
a Gaussian with a power-law tail. We fit the RS sample to determine the \Dz lifetime
and the time-resolution parameters, which are then held fixed in the fit to the WS sample.
We allow yields and background shape parameters to vary. Figure \ref{k3piFitFig}
shows the WS decay-time distribution and fit. Figure \ref{k3picont} shows 
contours of constant likelihood in the ($R_D$, $R_M$) plane; we find 
$R_M = (0.019^{+0.016}_{-0.015}\pm0.002)$\% assuming \CP conservation, and
$R_M = (0.017^{+0.017}_{-0.016}\pm0.003)$\% with \CP violation allowed. There
is no significant difference between the \Dz and \Dzb samples in the \CP-allowed
fit. 

\begin{table*}[!t]
\begin{tabular}{lccc}\hline
Mode                     & Luminosity [fb$^{-1}$] & Mixing                         & \CP violation \\ \hline\hline
\Dz\to\Km\pip            &  384                   & 3.9 $\sigma$                   & No evidence \\
\Dz\to\Km\Kp or \pip\pim &  384                   & 3.0 $\sigma$                   & No evidence \\
\Dz\to\Km\pip\piz        &  384                   & Exclude NM at 99\% CL          & No evidence \\ 
\Dz\to\Km\pip\pim\pip    &  230.4                 & Consistent with NM at 4.3\% CL & No evidence \\ 
\hline\hline
\end{tabular}
\caption{\label{restable} Summary of results.}
\end{table*}

To extract a consistency with the no-mixing hypothesis from these contours is
not quite straightforward, because the linear term in Equation \ref{k3pieqn} becomes
unconstrained as $R_M$ approaches zero. We therefore estimate the consistency of our
data with no-mixing using a frequentist method; we generate 1000 data sets of 76300
events each, setting the mixing parameters to zero in the generation. We then apply our
fit procedure to these sets; in 43 cases we find an $R_M$ equal to or greater than for
the data. We therefore conclude that our data are consistent with no-mixing only at the
4.3\% confidence level. 

We investigate systematic uncertainties from four sources, listed in order of 
decreasing significance. First is the $\sigma_t$ threshold, which we increase
from 0.5 to 0.6 ps. Second is the decay-time resolution function; we change this by
fixing one of the Gaussian widths to be exactly equal to $\sigma_t$, letting the
other constant of proportionality float as before. Third, the $m_{K3\pi}$ distribution
of the background is changed from exponential to a second-order polynomial. And fourth,
we use the nominal value of the \Dz lifetime instead of the one obtained from our
RS fit. Taken all together, these uncertainties are smaller than the statistical
uncertainty by a factor of five.

\section{Summary and outlook}

\babar\ has found evidence for mixing in several channels, as summarised
in Table \ref{restable}. With the total \babar\ luminosity expected to reach
750 fb$^{-1}$ before shutdown, or nearly twice the largest amount used in these
studies, we expect to be able to improve these measurements of the $D$ mixing parameters,
and to add other channels as well. 

\section{Acknowledgements}
We are grateful for the excellent luminosity and machine conditions
provided by our \pep2\ colleagues, 
and for the substantial dedicated effort from
the computing organizations that support \babar.
The collaborating institutions wish to thank 
SLAC for its support and kind hospitality. 
This work is supported by
DOE
and NSF (USA),
NSERC (Canada),
CEA and
CNRS-IN2P3
(France),
BMBF and DFG
(Germany),
INFN (Italy),
FOM (The Netherlands),
NFR (Norway),
MES (Russia),
MEC (Spain), and
STFC (United Kingdom). 
Individuals have received support from the
Marie Curie EIF (European Union) and
the A.~P.~Sloan Foundation.

\bibliographystyle{apsrev}

\end{document}